\documentclass[prb,twocolumn,showpacs,amsmath,amssymb,superscriptaddress]{revtex4-1}
\usepackage{graphicx}
\usepackage{dcolumn}
\usepackage{bm}
\usepackage{bbm}
\usepackage{ulem}
\usepackage{color}
\usepackage{slashed}
\usepackage{hyperref}
\usepackage{dsfont}

\newcommand{\nc}{\newcommand}

\nc{\be}{\begin{equation}} \nc{\ee}{\end{equation}}
\nc{\bea}{\begin{eqnarray}} \nc{\eea}{\end{eqnarray}}
\nc{\bean}{\begin{eqnarray*}} \nc{\eean}{\end{eqnarray*}}

\begin{document}

\title{Weyl fermions induced Magnon electrodynamics in Weyl semimetal}
\author{Jimmy A. Hutasoit}
\email{jah77@psu.edu}
\affiliation{Department of Physics, The Pennsylvania State University, University Park, Pennsylvania 16802}
\author{Jiadong Zang}
\affiliation{Department of Physics, John Hopkins University, Baltimore, Maryland 21218}
\author{Radu Roiban}
\affiliation{Department of Physics, The Pennsylvania State University, University Park, Pennsylvania 16802}
\author{Chao-Xing Liu}
\affiliation{Department of Physics, The Pennsylvania State University, University Park, Pennsylvania 16802}
\date{\today}

\begin{abstract}
Weyl fermions, which are fermions with definite chiralities, can give rise to anomalous breaking of the symmetry of the physical system which they are a part of. In their $(3+1)$-dimensional realizations in condensed matter systems, \textit{i.e.}, the so-called Weyl semimetals, this
anomaly gives rise to topological electromagnetic response of
magnetic fluctuations, which takes the form of non-local interaction
between magnetic fluctuations and electromagnetic fields. We study
the physical consequences of this non-local interaction, including
electric field assisted magnetization dynamics, an extra gapless
magnon dispersion, and polariton behaviors that feature ``sibling"
bands in small magnetic fields.
\end{abstract}

\pacs{}

\maketitle

In the 1980s, the study of anomalous behaviors of classically
conserved currents in systems with Weyl fermions revealed a deep
connection between this physical phenomena and the underlying
topology of the systems. In particular, it was realized that these
anomalies are deeply related to the skewness of the zero mode
structure of the Dirac operators, which in turn, using index
theorems, can then be related to characteristic classes, which are
topological invariants
\cite{Nielsen1977445,Nielsen1978475,alvarez1984topological,alvarez1985structure}.
Recently, with the advancement of realizations of topologically
ordered condensed matter systems, the interest on the connection between
topology and anomaly has been revived. Not only the study of
anomalies might give rise to a way to classify topological phases
in matters in the presence of interactions
\cite{PhysRevB.85.045104}, but it can also lead to topological
responses, which are physical manifestations of the underlying
topological nature, of these topologically ordered systems
\cite{PhysRevB.85.045104,PhysRevB.84.014527,PhysRevB.85.184503,PhysRevB.88.115307}.

In this letter, we study topological aspects of the Weyl semimetal,
a topologically protected semimetal with Weyl fermions. Weyl
semimetals can be regarded as a three-dimensional cousin of
graphene, where pairs of bands cross at certain points in the
momentum space, \textit{i.e}, the Weyl points. For a short
introduction to Weyl semimetals, see for example Ref.
\onlinecite{hosur2013recent}. Some material realizations of Weyl
semimetals consist of topological insulator heterostructures that
contain magnetic materials or magnetic dopants
\cite{Burkov:2011zr,cho2011possible}. An advantage of this
realization is that magnetic texture and fluctuations inherit some
physical properties that reflect the underlying topological nature
of this system. In particular, magnetic fluctuations are coupled to
Weyl fermions as an axial vector field \cite{Liu:2012ly} and
therefore, magnon excitations in this system possess topologically
non-trivial electromagnetic responses from the axial anomaly.

Our main result takes the form of a non-local interaction between
magnons and electromagnetic fields in Weyl semimetals, dictated by
the effective action Eq. (\ref{eq:eff}) below. The non-locality of
the interaction arises from the fact that the mediators of this
interaction are gapless excitations of Weyl fermions. The
modifications of the Landau-Lifshitz (LL) equation and Maxwell
equation due to this non-local interaction will give rise to two
physical consequences, which reflect the underlying topological
nature of Weyl semimetals. Firstly, in Weyl semimetals, electric fields can
couple to the local magnetic moments through gapless Weyl fermions, leading to an additional magnon excitation. 
Compared to the conventional spin wave in ferromagnet,
this new magnon branch is gapless and linear, inheriting the
nature of Weyl fermions. Secondly, the non-local
coupling between magnons and electromagnetic fields can induce a
magnon-polariton excitations in Weyl semimetals, which exhibit a
quite different spectrum from the usual polariton spectrum. In
particular, in small values of magnetic fields, there exists a band
with finite width that bifurcates into a pair of ``sibling" bands
with well-defined quasiparticles. 

Let us start by considering a topological insulator doped with magnetic impurities and assume that magnetic moments are magnetized along the growth direction, which we will take to be the $\hat{x}_3$-direction. This system can be realized in for example, Cr doped Bi$_2$Te$_3$ \cite{ADMA:ADMA201203493}. When magnetization is large enough, this model exhibits Weyl nodes, at which the effective excitations are two Weyl fermions with a relativistically-invariant dispersion relation. Thus, this system provides a natural description of Weyl semimetals using the 4-band model \cite{Liu:2012ly}, the details of which are given in Appx. \ref{appx:4}. It turns out that in this system, magnetic fluctuations of magnetic moments are coupled chirally to Weyl fermions \cite{Liu:2012ly}, and the effective action
describing the interaction between Weyl fermions, electromagnetic fields and magnetic fluctuations is given by
\bea
S = i \int d^4x  \, \bar{\psi} \gamma^{\mu} \left(\partial_{\mu} - i e A_{\mu} - i g \gamma_5 a_{\mu} \right) \psi,
\label{action}
\eea
where two Weyl fermions have been written together as a single Dirac fermion $\psi$,  $A_{\mu}$ is the electromagnetic
gauge field and $a_{\mu}$ is an axial vector field whose space-like components $\mathbf{a}$ are identified as
magnetic fluctuations \cite{Liu:2012ly}. Our convention for the $\gamma$ matrices 
\bea
\gamma^{\mu} =    \begin{pmatrix} 
      0 & \sigma^{\mu} \\
      \bar{\sigma}^{\mu} & 0 \\
   \end{pmatrix}; \quad\sigma^{\mu} = (\mathds{1}_{2\times 2}, \mathbf{\sigma}), \ \bar{\sigma}^{\mu} = (\mathds{1}_{2\times 2}, -\mathbf{\sigma}), \nonumber \\
\eea   
and the metric follows closely Ref. \onlinecite{Srednicki:2007quantum}, where the metric is mostly positive. In the following, we will consider only the case where the axial vector field strength $f_{\mu \nu} = \partial_{\mu} a_{\nu} - \partial_{\nu} a_{\mu}$ vanishes \footnote{When $f_{\mu \nu} =0$, the solution to Dirac equation is given by $\psi=0$. In this letter, we would like to obtain the effective interaction between magnons and electromagnetic fields by integrating out Weyl fermion fluctuations around the vacuum solution $\psi=0$. However, when the flux of the axial vector field strength takes non-zero integer values, the solution to Dirac equation consists of additional $(1+1)$-dimensional Weyl fermions \cite{Liu:2012ly}. The topological response obtained by integrating out fermionic fluctuations around this non-trivial background will be studied elsewhere.}, which is the case when there is no magnetic domain wall in the system \cite{Liu:2012ly}. Even though we are not going to use this fact here, it is worth noting for $f_{\mu \nu} =0$, the axial vector field can be written as $a_{\mu}=\partial_{\mu}\theta$, where $\theta$ has the physical meaning of axion fields \cite{PhysRevB.78.195424}.

To completely define this quantum field theory, it is necessary to specify a regularization scheme. This is particularly
important here as the chiral nature of the interactions \eqref{action} implies that the theory exhibits an anomaly
\cite{PhysRev.177.2426, bell1969pcac}, which appears as a  violation of current conservation in the three-point
function $\Gamma^{\mu\nu\rho}=\langle j^\mu(p)\, j^{\nu}(q) \,j^{\rho 5}(-p-q) \rangle$, where $j^\mu={\bar\psi}\gamma^\mu\psi$ and
$j^{\mu5}={\bar\psi}\gamma^\mu\gamma_5\psi$ are the $U(1)$ vector and axial current, respectively.
The anomaly is a reflection of the impossibility of simultaneously preserving the vector and axial symmetries
in the presence of any regulator. Since the vector symmetry characterizes the interaction of fermions and electromagnetic fields, the
correct definition of the theory must include a regularization scheme that respects the vector symmetry, which is nothing but the gauge invariance of electromagnetism. An example of such scheme is the dimensional regularization scheme of 't Hooft and Veltman \cite{t1972regularization}, and the calculation of $\Gamma^{\mu \nu\rho}$ using this scheme was done in Ref. \onlinecite{PhysRevD.20.3378}. One can also calculate this three-point function using Cutkosky rules and the dispersion relation, as was done in Ref. \onlinecite{hovrejvsi1992ultraviolet}. The result is
\be
\Gamma^{\mu\nu\rho}= - \frac{i e^2 g}{2 \pi^2} \, \varepsilon^{\alpha \mu \beta \nu} \, p_{\alpha} \, q_{\beta} \,\frac{g^{\rho \sigma} (p+q)_{\sigma}}{(p+q)^2}, \label{eq:3pt}
\ee
where $\varepsilon^{\alpha \beta \gamma \delta}$ is the totally antisymmetric Levi-Civita tensor.

It is easy to see that this three-point function satisfies the conservation of the vector current,
$p_\mu \, \Gamma^{\mu\nu\rho}=0 = q_\nu \, \Gamma^{\mu\nu\rho} $, but violates axial current conservation,
$(-p-q)_\rho \, \Gamma^{\mu\nu\rho} =  \frac{i e^2 g}{2 \pi^2} \, \varepsilon^{\alpha \mu \beta \nu} \, p_{\alpha} \, q_{\beta} \ne 0$.
We note that since anomalies are infrared phenomena (see for example, Ref. \onlinecite{Harvey:fk} and references within), we can expect the topological electromagnetic response of magnons to be insensitive to the details of the model away from the Weyl points as long as the electromagnetic gauge invariance is not broken. In a classic (particle physics) example, a similar anomaly is responsible for the decay of a neutral pion into two photons independently of the high energy completion of the theory of strong interactions that does not break the electromagnetic gauge invariance. For example, the pion decay is independent of the QCD quark masses~\cite{PhysRev.177.2426,bell1969pcac}. Nevertheless, it will be interesting to study the non-topological electromagnetic response of magnons from the high energy sector of Weyl semimetals and such study will be taken up elsewhere. For the rest of this letter, we will focus on studying the physical consequences of the anomalous term Eq. (\ref{eq:3pt}).

To that end, we construct the effective action of the topological electromagnetic response of magnons as follows
\bea
\label{eq:eff}
&&S_{\rm top} = \int \frac{d^4p}{(2\pi)^4} \frac{d^4q}{(2\pi)^4} \, \Gamma^{\mu\nu\rho}A_{\mu}(p) A_{\nu}(q) a_{\rho}(-p-q) \\
&=& -\frac{e^2 g}{8 \pi^2} \int d^4x \, d^4y \, \varepsilon^{\alpha \beta \gamma \delta} \, F_{\alpha \beta}(x) \, F_{\gamma \delta}(x) \, \frac{\partial G(x-y)}{\partial y^{\mu}} \, a^{\mu} (y), \nonumber
\eea
where $F_{\mu \nu} = \partial_{\mu} A_{\nu} - \partial_{\nu} A_{\mu}$ is the (vector) field strength, $G(x-y)$ is the Green function
of the d'Alembertian $\Box = \partial_{\mu} \partial^{\mu}$ and it obeys $\Box_x \, G(x-y) = \delta^4 (x-y)$.

We note that in the limit of a constant axial vector $a^\mu$ we recover the result of Refs. \cite{PhysRevB.86.115133,PhysRevB.87.161107}. For details, see Appx. \ref{appx:const}. Furthermore, using the definition $j^{\alpha} = \delta S/\delta A_{\alpha}$, we can obtain the anomalous Hall response
\be
j^{\alpha}(x) = -\frac{e^2 g}{2 \pi^2}\, \varepsilon^{\alpha \beta \gamma \delta} \, F_{\gamma \delta}(x)  \, \partial_{\beta}\int d^4y \, \frac{\partial G(x-y)}{\partial y^{\mu}} \, a^{\mu} (y),
\ee
which, in the limit of a constant axial vector $a^\mu$, reduces to the known result $j^{\alpha} = -\frac{e^2 g}{2 \pi^2}\, \varepsilon^{\alpha \beta \gamma \delta} a_{\beta}\,F_{\gamma \delta}$ of Ref. \onlinecite{PhysRevB.84.075129}.

As another non-trivial check, we can also compare the effective action Eq. (\ref{eq:eff}) with the result from 4-band model of Ref. \onlinecite{Liu:2012ly} at uniform magnetic field $\mathbf{B} = B \,\hat{x}_3$, akin to the calculation done in Ref. \onlinecite{PhysRevB.88.125105}. In this case, we have Landau levels and we can ask how the system responses to an applied electric field $\mathbf{E} = E \, \hat{x}_3$ and a perturbation due to magnetization. The result agrees with Eq. (\ref{eq:eff}) and for details, see Appx. \ref{appx:4}.

We are now ready to study the modifications of LL and Maxwell
equations caused by the topological response of Eq. (\ref{eq:eff}).
Assume an easy axis anisotropy is present such that the magnetic
moments are uniformly polarized along the $\hat{x}_3$ direction in
equilibrium. The magnon excitations are investigated by considering
the magnetization dynamics of the following Hamiltonian:
\be H_{\rm
magnet} = \frac{1}{2} \left(J \,(\mathbf{\nabla} \mathbf{M})^2 +
m^2\, | \mathbf{M}_{\parallel}|^2 \right) - \mathbf{B} \cdot
\mathbf{M},\label{eq:magnon1} \ee where $\mathbf{M}$ is the
magnetization, $\parallel$ denotes the in-plane direction, $\mathbf{B}$ is the magnetic field, and $m^2$ is the
easy axis anisotropy. Let $\mathbf{M} = M\, \hat{x}_3 +
\mathbf{a}$, with $M \gg a_i$. Substituting Eqs. (\ref{eq:eff}) and (\ref{eq:magnon1}) into the LL equation 
\be
\frac{d\hat{\mathbf{M}}}{dt}= \gamma \hat{\mathbf{M}}\times \frac{\partial (H_{\rm magnet} + H_{\rm top})}{\partial \hat{\mathbf{M}}},
\ee
we then have
\bea
\frac{\partial_t a_1}{\gamma M} &=& -  \frac{e^2 g}{8 \pi^2} \int d^4y \, \varepsilon^{\alpha \beta \gamma \delta} F_{\alpha \beta}(y) F_{\gamma \delta}(y)  \, \partial_2 G(x-y)\nonumber \\
& & +\, B_2 + \left(J \, \nabla^2 - m^2 - B_{3}/M \right)a_2, \label{eq:LL1} \\
\frac{\partial_t a_2}{\gamma M} &=& \frac{e^2 g}{8 \pi^2} \int d^4y \, \varepsilon^{\alpha \beta \gamma \delta} F_{\alpha \beta}(y) F_{\gamma \delta}(y)  \, \partial_1 G(x-y) \nonumber \\
& & - \,B_1 -  \left(J \, \nabla^2 - m^2 - B_{3}/M \right) a_1,
\label{eq:LL2} \eea where $\gamma$ is the product of the
gyromagnetic ratio, Bohr magneton and permeability of vacuum. It is
interesting to note that a spatial-dependent term contributed from
the Weyl fermion enters the magnetization dynamics. It plays the
same role as the in-plane magnetic field $B_1$ or $B_2$. The
magnetic moments experience this spatially modulated effective field
such that the magnon dispersion can be significantly changed. As the
electromagnetic field strength $F_{\alpha\beta}$ contains electric
fields as its component, it is quite interesting to see that the electric field can
dramatically modify magnetization dynamics. To illustrate this
more clearly, let us consider Weyl semimetals in a magnetic field along the $\hat{x}_3$ direction ${\bf B}=B_3\hat{x}_3$. 
For an oscillating electric field ${\mathbf E}=E_3 \,\exp\left( i \omega t- i {\mathbf{q}_{\parallel} \cdot \mathbf{x}_{\parallel}}\right)\,\hat{x}_3$, we obtain
\bea
\mathbf{a}_{\parallel}(\omega, \mathbf{q}_{\parallel}) &=& -\frac{ie^2 g}{\pi^2} \,\frac{ (J \,q^2+m^2+ B_{3}/M) \,\mathbf{q}_{\parallel}}{\omega^2/(\gamma^2 M^2) - (J\, q^2+m^2+ B_{3}/M)^2}\nonumber \\
& &  \frac{E_3 \,B_3}{\omega^2-v_F^2 \,q^2}, \label{eq:magnon} \eea
where $q = |\mathbf{q}_{\parallel}|$. Here, we have inserted back the Fermi velocity of the Weyl fermions
in order to differentiate it with the speed of light in the medium,
which we are taking to be unity.

Two poles of $\mathbf{a}_{\parallel}(\omega, \mathbf{q}_{\parallel})$ suggest the
existence of two magnon branches in this system. In addition to the
usual spin wave $\omega = \gamma \,M\,(J \,q^2+m^2+ B_{3}/M)$,
a magnon with $\omega = v_F\,|q|$ is present. This novel dispersion
is determined solely by a property of the Weyl fermions, namely
their Fermi velocity. More importantly, this new branch is gapless,
leading to a long-range correlation of spin excitations. 
Physically, this magnon excitation can be understood as a direct result of the coupling between two magnetic moments mediated by Weyl fermions. The gapless nature of Weyl fermions leads to long-range correlation of this magnon excitation. 
Therefore, this new magnon
dispersion is a distinct feature of Weyl semimetals. One can then 
employ neutron scattering experiments to test our prediction.


Let us now look at the Maxwell equation in the presence of topological response of Eq. (\ref{eq:eff}). It is given by
\bea
- \partial_{\alpha}F^{\alpha \beta} - \frac{e^2 g}{2 \pi^2} \,\varepsilon^{\alpha \beta \gamma \delta}F_{\gamma \delta} \int d^4y \, \frac{\partial G(x-y)}{\partial x^{\alpha}} \, \frac{\partial a^{\mu}(y)}{\partial y^{\mu}} &=& 0. \nonumber \\ \label{eq:eomA}
\eea
By using the identity $\varepsilon_{\mu \nu \rho \sigma} \partial^{\rho} F^{\mu \nu} = 0$ and keeping only the linear term in $\mathbf{E}$, we get
\bea
- \partial^2_t \mathbf{E} + \nabla^2 \mathbf{E} - \frac{e^2g}{\pi^2}\, \mathbf{B} \int d^4y \, \partial^2_t G(x-y) \, \left(\mathbf{\nabla} \cdot \mathbf{a}\right)(y) \nonumber \\
+ \frac{e^2g}{\pi^2}\, \mathbf{\nabla}\left(\mathbf{B} \cdot \mathbf{\nabla} \right) \int d^4y \,  G(x-y) \, \left(\mathbf{\nabla} \cdot \mathbf{a}\right)(y) &=& 0. \nonumber \\ \label{eq:eomlinearE}
\eea
For concreteness, let us again consider applying a uniform magnetic field $B_3$ along the $\hat{x}_3$ direction. If we shine a light with the electric field 
${\mathbf E}=E_3 \,\exp\left( i \omega t- i {\mathbf{q}_{\parallel} \cdot \mathbf{x}_{\parallel}}\right)\,\hat{x}_3$, we have 
\bea
&&\left[\omega^2 - q^2 - \left(\frac{e^2 g\, B_3}{\pi^2}\right)^2 \frac{q^2 \, \omega^2}{(\omega^2-v_F^2\,q^2)^2} \right. \nonumber \\
&& \ \ \left. \frac{J\,q^2+m^2+ B_{3}/M}{\omega^2/(\gamma^2 M^2) - (J\, q^2+m^2+ B_{3}/M)^2} \right] E_3(\omega,\mathbf{q})=0. \nonumber \\ \label{eq:polariton}
\eea

\begin{figure*} 
   \centering
\begin{tabular}{l}
   \includegraphics[width=6.5in]{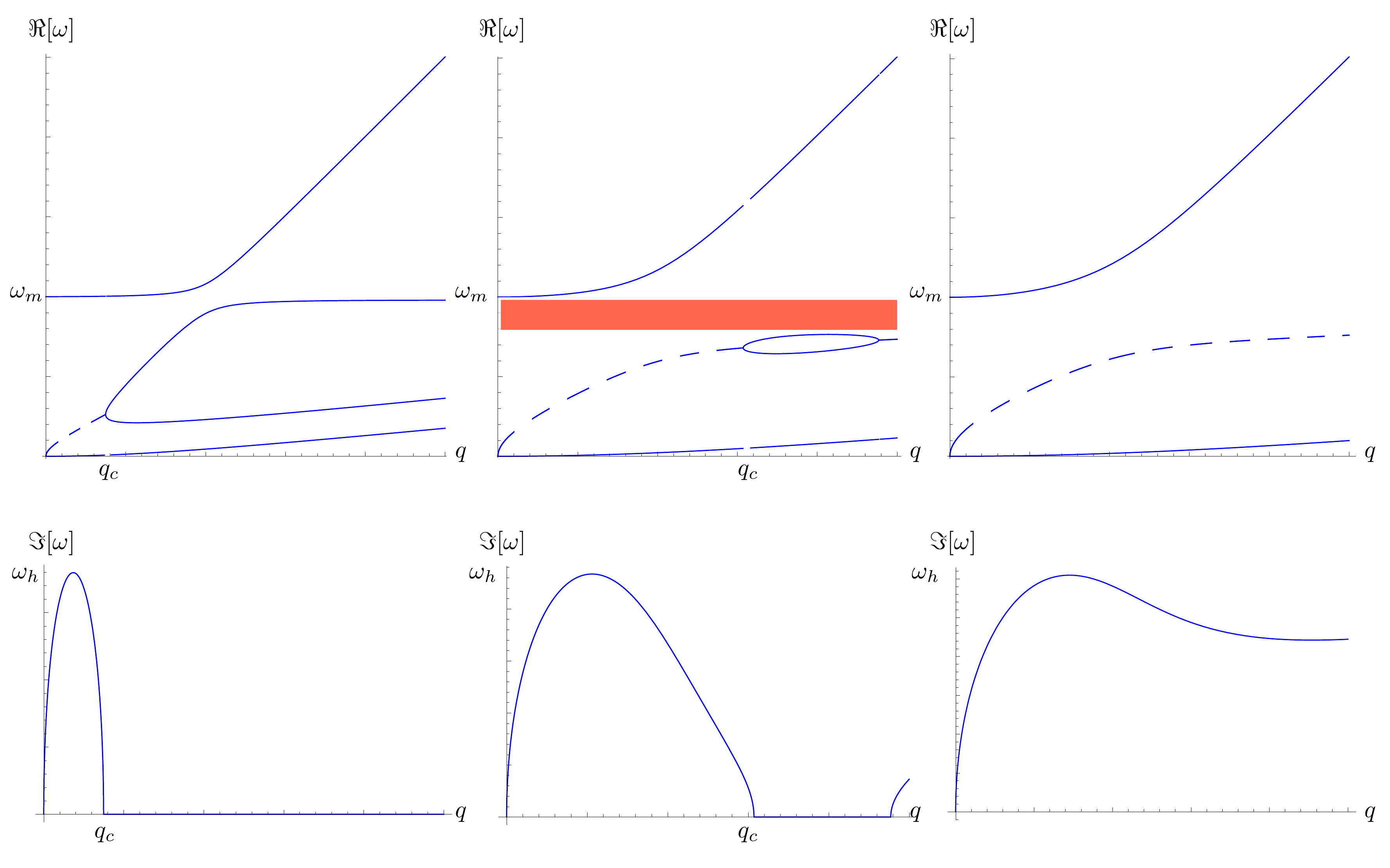}
\end{tabular}
   \caption{The polariton spectra as we increase the magnetic field from left to right. Top: Real parts of the polariton poles, where the solid lines correspond to those with vanishing imaginary part while the dashed lines otherwise. Here, the magnon gap is given by $\omega_m=\gamma (M m^2+B_3)$ and the bifurcation starts at $q_c \approx 2 e g B_3 /\left[\pi \gamma (M m+\sqrt{M B_3})\right]$, which we estimate to be $\sim 10^{-5} - 10^{-4}$ A$^{-1}$. We have also shown the forbidden band for the intermediate value of magnetic field, depicted as the shaded region. Bottom: Imaginary part for the intermediate band, whose maximum value is given by $\omega_h$, with $\omega_h \approx\sqrt{2} eg B_3/\left[\pi \gamma (M m+\sqrt{M B_3})\right]$.}
   \label{fig:spec}
\end{figure*}

The solutions of the equation above correspond to the poles of the polariton modes, which can then be detected using various spectroscopy techniques, such as angle-resolved electron energy-loss spectroscopy. In a typical magnetic material, in the long wavelength regime, the effective magnon velocity $2 \gamma M J q$ is significantly smaller than the Fermi velocity of the Weyl fermions. Thus, the dispersion of the magnon can be neglected. The typical behavior of the real part of the poles are plotted in Fig.  \ref{fig:spec}. See also Appx. \ref{appx:green}.

One can find four bands in total in Fig. \ref{fig:spec}. These solutions represent the hybridization between electric fields and magnetic moments due to the non-local coupling induced by Weyl fermions. At a non-vanishing magnetic field, the top and bottom bands are non-degenerate. Furthermore, the imaginary parts of their respective poles vanish and therefore, their spectral density is given by a Dirac $\delta$-function. The intermediate band, however, acquires a non-vanishing imaginary part of its pole and therefore, feature broadened spectral density. This broadening is due to its ability to emit Weyl fermions, which results in it acquiring complex self-energy. This is not unlike the physics of plasmon, see for example Ref. \onlinecite{Mahan}. At a low magnetic field, this band bifurcate into a pair of ``sibling" bands, whose spectral densities are given by Dirac $\delta$-functions, where the threshold for emitting Weyl fermions is beyond the energetics. As the magnetic field increases, this bifurcation disappears. The value of magnetic field at which this happens scales as $g^{-2}$ and for $g \sim 0.1$, this value is of order 10 T.

At a small $q$, the top and bottom bands scale as $\omega \sim q^0$ and $q^2$, respectively, while the intermediate band scales like $\omega \sim q^{1/2}$. 
We note that for the intermediate band, 
there is a regime where the velocity of the latter exceeds the speed of light in the Weyl semimetal. This ``tachyonic" regime needs to be excised, similar to the case of surface optical phonon for a polar crystal such as NaCl \cite{mahan2010condensed}.

The polariton spectrum also features an energy range at which there exists no polariton modes. This ``forbidden" band is particularly manifest at larger values of magnetic field. Therefore, the incident light will be totally reflected if its frequency lies within the forbidden band. Such forbidden band is predicted to be a generic feature of topological magnetic insulator \cite{li2010dynamical}, however, sibling bands are particular to the Weyl semimetal.

In order to probe the polariton, it is crucial that the energy dumped into the system is spent to excite the polariton and not the Weyl
fermions. In other words, the observability of the polariton
spectrum depends heavily on how much it overlaps with the single
particle excitation regimes of the Weyl fermions. We find indeed
that this overlap is negligible as the typical minimum energy needed
to excite the Weyl fermions is about $10$ meV (see Appx. \ref{appx:4} for
details) while the typical magnon gap $\omega_m=\gamma (M\,m^2+
B_{3})$ is about $0.1$ meV \cite{popova2001structure}.

\textit{Summary} -- In this article, we have shown that the topological response of magnons in Weyl semimetal is given by a non-local interaction between magnons and electromagnetic fields. This non-local interaction manifests itself in term of electric-field-induced magnetization dynamics that results in gapless magnon excitations. It also gives rise to resonant behavior in the form of magnon polariton featuring sibling bands and forbidden band.

\textit{Acknowledgements} -- We would like to thank Gerald Mahan, Xiaoliang Qi, Cenke Xu and
Jainendra Jain for insightful discussions. J. H. is supported by NSF
grant DMR-1005536 and DMR-0820404 (Penn State MRSEC). J. Z. is
supported by the Theoretical Interdisciplinary Physics and
Astrophysics Center and by the U.S. Department of Energy, Office of
Basic Energy Sciences, Division of Materials Sciences and
Engineering under Award DEFG02-08ER46544. R. R. is supported by the
U.S. Department of Energy under contract DE-SC0008745.

\appendix
\begin{widetext}
\section{4-band model calculations \label{appx:4}}

Let us start with the 4-band model of Ref. \onlinecite{Liu:2012ly}
\be
H = H_0 + H_1,
\ee
where
\bea
\label{4band}
H_0 =   \left( \begin{matrix} 
      {\cal M} +M & 0 & - i \,L_1 \,k_3 & i\, L_2\, k_- \\
      0 & {\cal M} - M & - i\, L_2 \,k_+ & - i \,L_1 \,k_3  \\
      i\, L_1\, k_3 & i\, L_2 \,k_- & - {\cal M} + M & 0 \\
      - i \, L_2 \, k_+ & i \, L_1 \, k_3 & 0 & - {\cal M} - M
   \end{matrix} \right)\ ,~~~ \label{eq:4band}
\eea
with
\bea
{\cal M} &=& M_0 + M_1 \, k_3^2 + M_2 \, k_{\parallel}^2,  \\
k_{\pm} &=& k_1 \pm i \, k_2 ;
\eea
and 
\be
H_1 = {\rm diag}(\tilde{g} \mu_3,-\tilde{g} \mu_3,\tilde{g} \mu_3,-\tilde{g} \mu_3) = \tilde{g} \mu_3 \sigma_3  \cdot \mathds{1}_{2 \times 2}.
\ee
Here, we have magnetized the system along the $\hat{x}_3$-direction with magnetization $M$ and for simplicity, allow magnetic fluctuations only along that same direction, where $\tilde{g} \ll 1$. All the material related parameters are defined in Ref. \onlinecite{Liu:2012ly}. 

For $|M| > |M_0|$, this model exhibits Weyl points. Expanding around these Weyl points, one can obtain the low energy effective theory of Weyl fermions coupled chirally to the magnetic fluctuations as in Eq. (\ref{action}). In particular, the axial vector field can then be related to the magnetic fluctuations as
\be
a_3 =  - \frac{M}{L_1 \sqrt{M^2 - M_0^2}} \mu_3.
\ee
For details, see Ref. \onlinecite{Liu:2012ly}.

Let us now turn on the external uniform magnetic field $\mathbf{B} = B \,\hat{x}_3$. The conjugate momenta on the direction perpendicular to the magnetic field become $k_+ \rightarrow \Pi_+ = \sqrt{2} \,a/ \ell_c$ and $k_- \rightarrow \Pi_- = \sqrt{2} \,a^{\dagger}/ \ell_c$, where $a$ and $a^{\dagger}$ are the annihilation and creation operators for the Landau levels, respectively, and $\ell_c$ is the magnetic length. Since $a \, \phi_n = \sqrt{n} \, \phi_{n-1}$ and $a^{\dagger} \, \phi_{n} = \sqrt{n+1} \, \phi_{n+1}$, writing the wave function as
\be
\phi =    \left(\begin{matrix} 
      f_1^n \, \phi_{n-1} \\ 
      f_2^n \, \phi_n\\
      f_3^n \, \phi_{n-1} \\
      f_4^n \, \phi_n \\
   \end{matrix} \right),
\ee
the Hamiltonian then can be written as
\bea
H_{\rm LL} &=& \left(   \begin{matrix} 
      {\cal M}_n+M & 0 & - i \,L_1 \,k_3 & \frac{i \, \sqrt{2} L_2}{\ell_c}\sqrt{n}\\
      0 &  {\cal M}_n-M& - \frac{i \, \sqrt{2} L_2}{\ell_c}\sqrt{n} & - i \,L_1 \,k_3  \\
       i \,L_1 \,k_3 & \frac{i \, \sqrt{2} L_2}{\ell_c}\sqrt{n} & - {\cal M}_n+M & 0 \\
       - \frac{i \, \sqrt{2} L_2}{\ell_c}\sqrt{n} & i \,L_1 \,k_3 & 0 & - {\cal M}_n-M
          \end{matrix} \right),
\eea
where
\be
{\cal M}_n = M_0+M_1\, k_3^2 + \frac{2 M_2}{\ell_c^2} \left(n-\frac{1}{2}\right).
\ee
Diagonalizing this Hamiltonian, we can then obtain the Landau levels for $n>0$. For $n=0$, this Hamiltonian is reduced to half in size
\bea
H_{\rm LLL} &=& \left(   \begin{matrix} 
       {\cal M}_0-M & - i \,L_1 \,k_3  \\
      i \,L_1 \,k_3 & - {\cal M}_0-M
          \end{matrix} \right).
\eea
The resulting Landau levels are plotted in Fig. \ref{fig:landau}.

\begin{figure}[h] 
   \centering
   \includegraphics[width=5in]{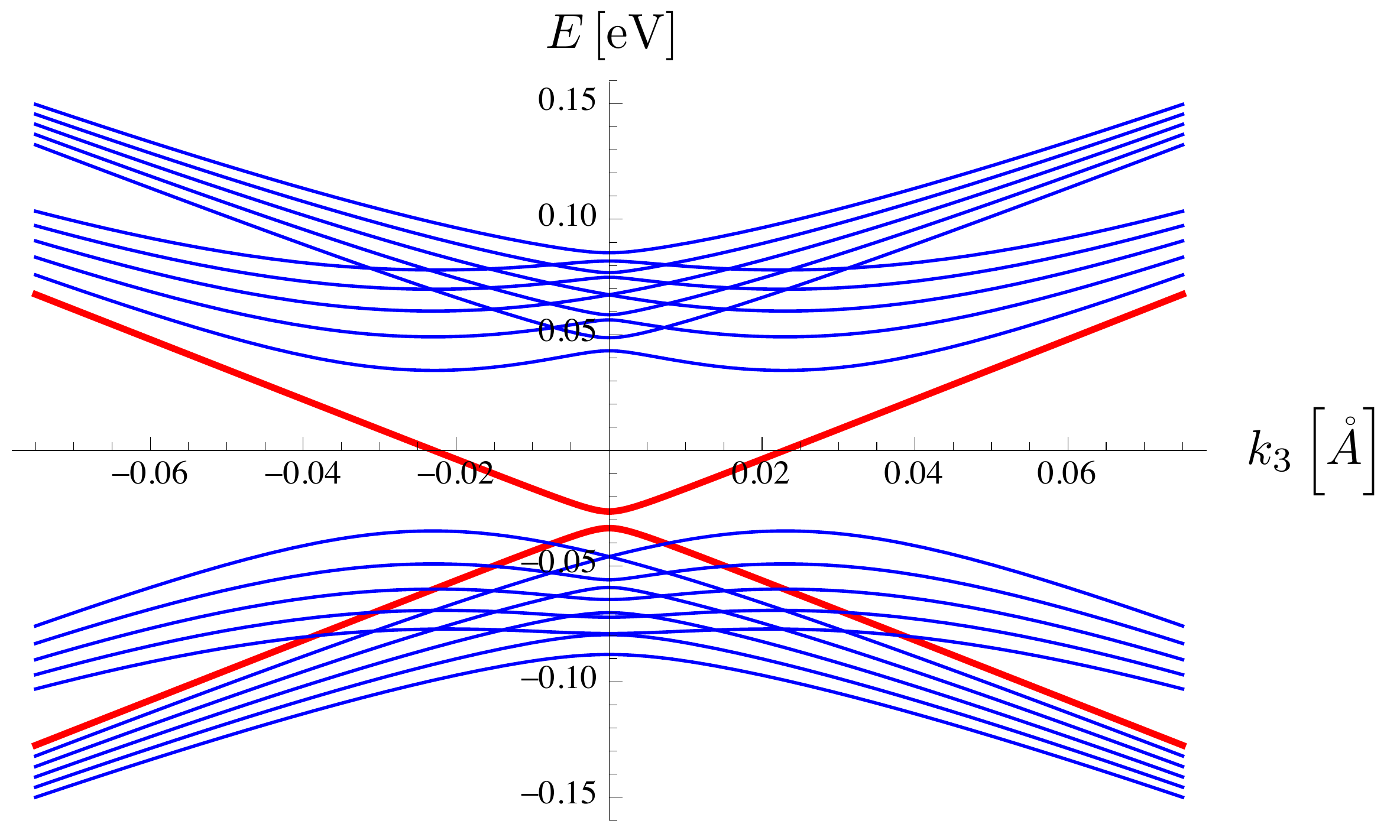} 
   \caption{Landau levels for $M_0 = -0.005$ eV, $M_1 = 0.342$ eV$\cdot$$\mathring{\rm A}^2$, $M_2 = 18.225$ eV$\cdot$$\mathring{\rm A}^2$, $L_1 = 1.3$ eV$\cdot$$\mathring{\rm A}$, $L_2 = 2.82$ eV$\cdot$$\mathring{\rm A}^2$ and $B = 5$T. $n = 0$ Landau levels are depicted in red, while the higher Landau levels are depicted in blue.}
   \label{fig:landau}
\end{figure}

We can now perturb the above Hamiltonian by applying an external electric field $\mathbf{E} = E \, \hat{x}_3$ and ask what the response of the system to the axial vector field is. The response function is given by
\bea
\Pi_{aE} (\omega,q \hat{x}_3) &=& \frac{\delta^2 S}{\delta a_3 \, \delta E} \nonumber \\
&=& \frac{eB}{2\pi \ell_3} \sum_{n,n'\in LL} \sum_{k_3} \frac{n_F[\varepsilon_n(k_3)] - n_F\left[\varepsilon_{n'}(k_3 + q)\right]}{\omega + \varepsilon_n(k_3) - \varepsilon_{n'}(k_3 + q)} \, (i\, \tilde{g} \,e)\,\langle n',k_3+q|   \mathbb{I}_{4\times 4} | n, k_3 \rangle \nonumber \\
& & \qquad\qquad\qquad  \left(- \frac{L_1 \sqrt{M^2 - M_0^2}}{M} \right)\langle n, k_3 \left| \sigma_3 \cdot \mathbb{I}_{2 \times 2} \right| n',k_3+q \rangle.
\eea
Here, $n_F$ is Fermi distribution, $\ell_3$ is the length of the system in $\hat{x}_3$ direction and the factor $e\,B$ comes from the degeneracy of Landau levels. We can approximate this by neglecting the contribution from higher Landau levels 
\bea
\Pi_{aE} (\omega,q\, \hat{x}_3) &=& \frac{L_1 \sqrt{M^2 - M_0^2}}{M} \, \frac{i \,\tilde{g} \,e^2\, B}{2\pi \ell_3} \sum_{n,n'\in {\rm LLL}} \sum_{k_3} \frac{n_F[\varepsilon_n(k_3)] - n_F\left[\varepsilon_{n'}(k_3 + q)\right]}{\omega + \varepsilon_n(k_3) - \varepsilon_{n'}(k_3 + q)} \, \Big| \langle n',k_3+q| n, k_3 \rangle \Big|^2, \nonumber \\
\eea
where we have projected the matrix elements of the interaction Hamiltonian $H_1$ (which is $4\times4$) into the LLL space (which is $2\times2$). For $T=0$, $\mu = 0$ and small $q$, we therefore have
\bea
\Pi_{aE}  (\omega\ll q,q\, \hat{x}_3) &\approx& \left[L_1^4+4 L_1^2 M_0 M_1 + 4 M_1^2 M^2 - (2 M_0 M_1 + L_1^2)\sqrt{L_1^4+4 L_1^2 M_0 M_1 + 4 M_1^2 M^2} \, \right]^{-\frac{1}{2}}\nonumber \\
& &\sqrt{2} \, \pi \,|M_1|\,L_1 \sqrt{M^2 - M_0^2} \, \frac{\tilde{g} \,e^2\, B}{\pi^2\, (- i \, q)}\nonumber \\
&\equiv& \frac{g \,e^2\, B}{\pi^2\,} \left(\frac{- i \, q}{-q^2}\right). \label{eq:response}
\eea
The Lagrangian density in momentum space is then given by
\bea
{\cal L} = \Pi_{a E}\, E \,a_3 =  \frac{g \,e^2}{\pi^2\,} \,E\,B \left(\frac{- i \, q}{-q^2}\right) a_3,
\eea
which upon Fourier transforming back to real space, reads
\bea
{\cal L} =  - \frac{g \,e^2}{\pi^2\,} \,E(x)\,B(x) \, \mathbf{\nabla}_y G(x-y) \cdot \mathbf{a}(y),
\eea
in agreement with Eq. (\ref{eq:eff}).

Next, let us look at the polarization operator in the presence of the external magnetic field. The regime where its imaginary part is non-vanishing corresponds to the regime of single particle excitations (SPE) of the Weyl fermions. Since we are interested in comparing it to the spectrum of the polariton, we are going to focus on the case where the momentum is perpendicular to the direction of the magnetic field $\mathbf{q} = \mathbf{q}_{\parallel}$. The polarization operator is then given by 
\bea
\Pi_{EE} (\omega,\mathbf{q}_{\parallel}) &=& \frac{\delta^2 S}{ (\delta E)^2} \nonumber \\
&=& \frac{e^3\, B}{2\pi \ell_3} \sum_{n,n'} \sum_{k_3} \frac{n_F[\varepsilon_n(k_3)] - n_F\left[\varepsilon_{n'}(k_3)\right]}{\omega + \varepsilon_n(k_3) - \varepsilon_{n'}(k_3)+i 0^+} \, \Big| \langle n',k_3| n, k_3 \rangle \Big|^2.
\eea
We note that the right hand side does not depend on $\mathbf{q}_{\parallel}$. Furthermore, the bottom boundary of the SPE regime is the smallest gap between the filled part of the lowest Landau level and the second Landau level. As can be seen from Fig. \ref{fig:landau}, it is of order $0.03$ eV.

\section{The Constant Vector Limit \label{appx:const}}

Let us start by putting our theory, Eq. (\ref{eq:eff}), in a finite volume by introducing a finite volume regulator $a^{\mu}(y) \rightarrow a^{\mu} (y) \exp\left[-|\hat{a}\cdot y|/\Lambda\right]$, such that
\bea
S_{\rm top} =-\frac{e^2 g}{8 \pi^2} \int d^4x \, d^4y \, \varepsilon^{\alpha \beta \gamma \delta} \, F_{\alpha \beta}(x) \, F_{\gamma \delta}(x) \, \frac{\partial G(x-y)}{\partial y^{\mu}} \, a^{\mu} (y) e^{-\frac{|\hat{a}\cdot y|}{\Lambda}}, \label{eq:eff'}
\eea
where $\Lambda \gg L$ and $L$ is the typical size of the system. We note that when the axial vector field goes to a constant vector limit, the field strength $\partial_{\mu} \left[a_{\nu} (y) \exp\left(-|\hat{a}\cdot y|/\Lambda\right) \right]- \partial_{\nu} \left[ a_{\mu} (y) \exp\left(-|\hat{a}\cdot y|/\Lambda\right)\right]$, which includes the curl $\mathbf{\nabla} \times [\mathbf{a}\, \exp(-|\hat{a}\cdot y|/\Lambda)]$, remains vanishing while the divergence $\partial_{\mu} [a^{\mu}\, \exp(-|\hat{a}\cdot y|/\Lambda)]$ remains non-zero. Therefore, even at the constant vector limit, the magnon is Helmholtz decomposed into the curl-free term only.

In order to obtain the constant vector limit of Eq. (\ref{eq:eff'}), we write $a^\mu \exp(-|\hat{a}\cdot y|/\Lambda) = g^{\mu \nu} \, \partial_\nu \left[a\cdot y\, \exp(-|\hat{a}\cdot y|/\Lambda)\right] + {\cal O}(1/\Lambda)$. Substituting it in Eq.~\eqref{eq:eff} and integrating by parts, we obtain
\bea
S_{\rm top} &=& - \frac{e^2 g}{8 \pi^2} \int d^4x \, d^4y \, \varepsilon^{\alpha \beta \gamma \delta} \, F_{\alpha \beta}(x) \, F_{\gamma \delta}(x) \, \partial_{\nu}\left[g^{\mu \nu}\frac{\partial G(x-y)}{\partial y^{\mu}} \, a\cdot y \,e^{-\frac{|\hat{a}\cdot y|}{\Lambda}}\right] \nonumber \\
& & \,+ \frac{e^2 g}{8 \pi^2} \int d^4x \, d^4y \, \varepsilon^{\alpha \beta \gamma \delta} \, F_{\alpha \beta}(x) \, F_{\gamma \delta}(x) \, \Box_y G(x-y) \, a\cdot y \,e^{-\frac{|\hat{a}\cdot y|}{\Lambda}}.
\eea
The first term is the surface term that vanishes due to the regulator and using the definition of the Green's function we recover the action in the Ref. \onlinecite{PhysRevB.86.115133}
\bea
S_{\rm top} \rightarrow\frac{e^2 g}{8 \pi^2} \int d^4x \, \varepsilon^{\alpha \beta \gamma \delta} \, F_{\alpha \beta}(x) \, F_{\gamma \delta}(x) \, (a\cdot x).
\eea

\section{The Polariton Green Function\label{appx:green}}
\begin{figure}[h] 
   \centering
\begin{tabular}{l}
   \includegraphics[width=5in]{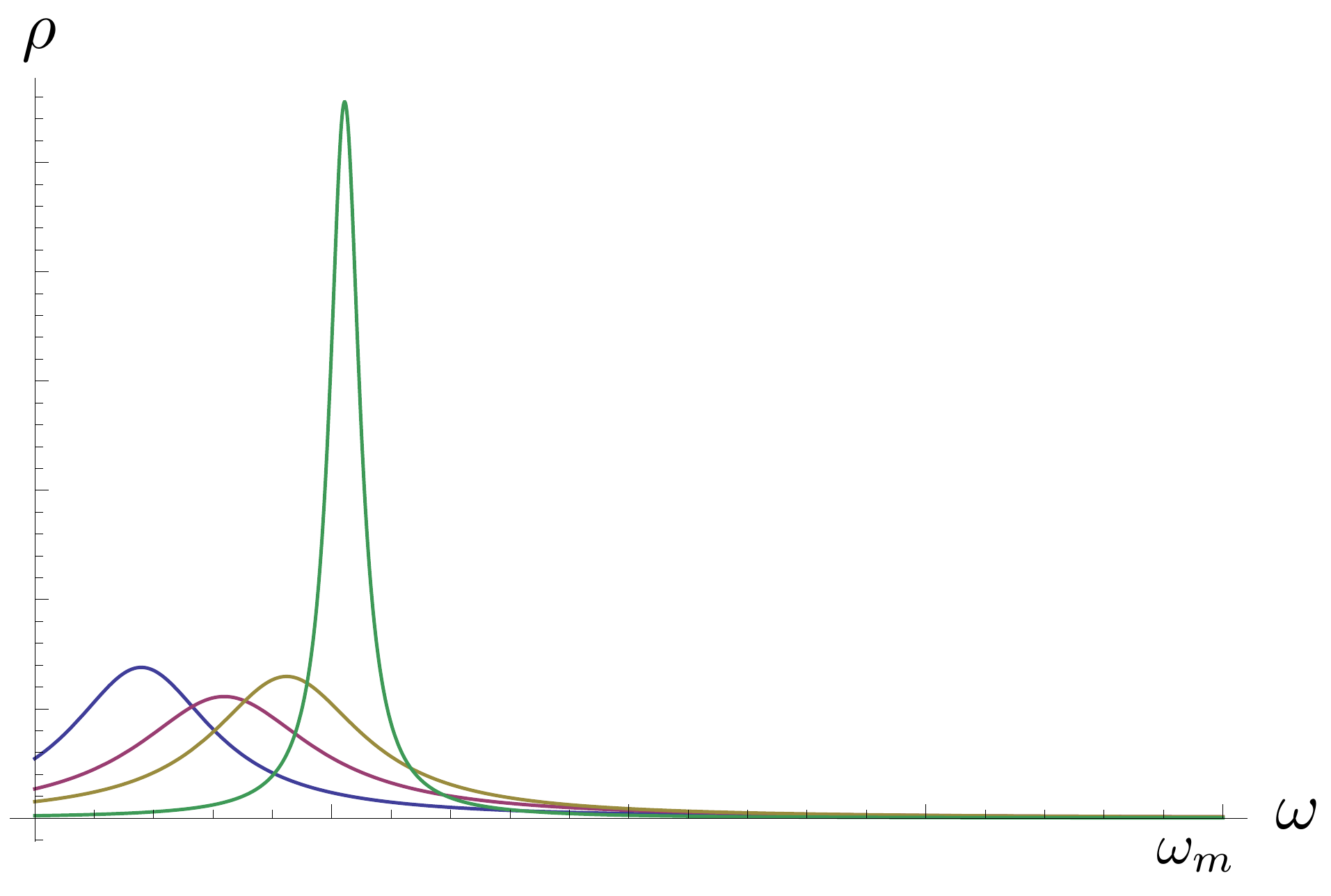} 
\end{tabular}
   \caption{The spectral density at small magnetic field for different values of momenta $q/q_c =  0.2$, $0.5$, $0.75$ and $0.99$ for the blue, purple, yellow and green lines (from left to right), respectively. Here, $q_c \approx 2 e g B_3 /\left[\pi \gamma (M m+\sqrt{M B_3})\right]$ is the momentum at which the bifurcation into the ``sibling" bands starts and $\omega_m=\gamma (M m^2+B_3)$ is the magnon gap.}
   \label{fig:dens}
\end{figure}

Setting ${\mathbf B} = B \hat{x}_3$ and then applying an electric field along the $\hat{x}_3$ direction, the Landau-Lifshitz and Maxwell equations can be written as 
\be 
{\cal E}(\omega,q)  \begin{pmatrix} 
      \frac{\mathbf{q}_{\parallel} \cdot \mathbf{a}_{\parallel}}{q} \\
      E_3\\
   \end{pmatrix} = 0, \label{eq:eom}
\ee
where $|\mathbf{q}_{\parallel}| = q$ and 
\be
 {\cal E}(\omega,q) =   \begin{pmatrix} 
       \frac{\omega^2}{\gamma^2 M^2} - \left(J\, q^2+m^2 + \frac{B_3}{M}\right)^2 & \frac{e^2 g}{\pi^2} \frac{iq(J \,q^2+m^2+B_3/M) B_3}{\omega^2-v_F^2 \,q^2}   \\
     - \frac{e^2 g}{\pi^2} \frac{i \, q \, \omega^2 \, B_3}{\omega^2-v_F^2\,q^2} & \omega^2 - q^2 \\
   \end{pmatrix}.
\ee   
The Green function for the polariton then must satisfy
\be
 {\cal E}(\omega,q)\,  {\cal G}(\omega,q) = \mathbb{I}_{2\times 2},
 \ee 
and its singularities are given by the singularities of ${\cal E}^{-1}$, which are the zeroes of $\det {\cal E}$. We note that the solutions to $\det {\cal E}=0$ are identical to the solutions of Eq. (\ref{eq:polariton}). Furthermore, we can obtain the spectral density from $\rho (\omega, q) = \Im[ {\cal G} (\omega,q)]$. The spectral densities of the top and bottom band are trivial as they correspond to well-defined quasiparticles, while the spectral density of the intermediate band exhibits finite width. In Fig. \ref{fig:dens}, we plot the spectral density of the intermediate band at small magnetic field for different values of momenta up to $q=0.99 q_c$, where $q_c$ is the momentum at which bifurcation into the ``sibling" bands occur.
\end{widetext}

\bibliography{References}

\begin{thebibliography}{30}%
\makeatletter
\providecommand \@ifxundefined [1]{%
 \@ifx{#1\undefined}
}%
\providecommand \@ifnum [1]{%
 \ifnum #1\expandafter \@firstoftwo
 \else \expandafter \@secondoftwo
 \fi
}%
\providecommand \@ifx [1]{%
 \ifx #1\expandafter \@firstoftwo
 \else \expandafter \@secondoftwo
 \fi
}%
\providecommand \natexlab [1]{#1}%
\providecommand \enquote  [1]{``#1''}%
\providecommand \bibnamefont  [1]{#1}%
\providecommand \bibfnamefont [1]{#1}%
\providecommand \citenamefont [1]{#1}%
\providecommand \href@noop [0]{\@secondoftwo}%
\providecommand \href [0]{\begingroup \@sanitize@url \@href}%
\providecommand \@href[1]{\@@startlink{#1}\@@href}%
\providecommand \@@href[1]{\endgroup#1\@@endlink}%
\providecommand \@sanitize@url [0]{\catcode `\\12\catcode `\$12\catcode
  `\&12\catcode `\#12\catcode `\^12\catcode `\_12\catcode `\%12\relax}%
\providecommand \@@startlink[1]{}%
\providecommand \@@endlink[0]{}%
\providecommand \url  [0]{\begingroup\@sanitize@url \@url }%
\providecommand \@url [1]{\endgroup\@href {#1}{\urlprefix }}%
\providecommand \urlprefix  [0]{URL }%
\providecommand \Eprint [0]{\href }%
\providecommand \doibase [0]{http://dx.doi.org/}%
\providecommand \selectlanguage [0]{\@gobble}%
\providecommand \bibinfo  [0]{\@secondoftwo}%
\providecommand \bibfield  [0]{\@secondoftwo}%
\providecommand \translation [1]{[#1]}%
\providecommand \BibitemOpen [0]{}%
\providecommand \bibitemStop [0]{}%
\providecommand \bibitemNoStop [0]{.\EOS\space}%
\providecommand \EOS [0]{\spacefactor3000\relax}%
\providecommand \BibitemShut  [1]{\csname bibitem#1\endcsname}%
\let\auto@bib@innerbib\@empty
\bibitem [{\citenamefont {Nielsen}\ \emph {et~al.}(1977)\citenamefont
  {Nielsen}, \citenamefont {Romer},\ and\ \citenamefont
  {Schroer}}]{Nielsen1977445}%
  \BibitemOpen
  \bibfield  {author} {\bibinfo {author} {\bibfnamefont {N.}~\bibnamefont
  {Nielsen}}, \bibinfo {author} {\bibfnamefont {H.}~\bibnamefont {Romer}}, \
  and\ \bibinfo {author} {\bibfnamefont {B.}~\bibnamefont {Schroer}},\ }\href
  {http://www.sciencedirect.com/science/article/pii/0370269377904105}
  {\bibfield  {journal} {\bibinfo  {journal} {Phys. Lett. B}\ }\textbf
  {\bibinfo {volume} {70}},\ \bibinfo {pages} {445 } (\bibinfo {year}
  {1977})}\BibitemShut {NoStop}%
\bibitem [{\citenamefont {Nielsen}\ \emph {et~al.}(1978)\citenamefont
  {Nielsen}, \citenamefont {Romer},\ and\ \citenamefont
  {Schroer}}]{Nielsen1978475}%
  \BibitemOpen
  \bibfield  {author} {\bibinfo {author} {\bibfnamefont {N.}~\bibnamefont
  {Nielsen}}, \bibinfo {author} {\bibfnamefont {H.}~\bibnamefont {Romer}}, \
  and\ \bibinfo {author} {\bibfnamefont {B.}~\bibnamefont {Schroer}},\ }\href
  {http://www.sciencedirect.com/science/article/pii/0550321378902717}
  {\bibfield  {journal} {\bibinfo  {journal} {Nucl. Phys. B}\ }\textbf
  {\bibinfo {volume} {136}},\ \bibinfo {pages} {475 } (\bibinfo {year}
  {1978})}\BibitemShut {NoStop}%
\bibitem [{\citenamefont {Alvarez-Gaume}\ and\ \citenamefont
  {Ginsparg}(1984)}]{alvarez1984topological}%
  \BibitemOpen
  \bibfield  {author} {\bibinfo {author} {\bibfnamefont {L.}~\bibnamefont
  {Alvarez-Gaume}}\ and\ \bibinfo {author} {\bibfnamefont {P.}~\bibnamefont
  {Ginsparg}},\ }\href@noop {} {\bibfield  {journal} {\bibinfo  {journal}
  {Nucl. Phys. B}\ }\textbf {\bibinfo {volume} {243}},\ \bibinfo {pages} {449}
  (\bibinfo {year} {1984})}\BibitemShut {NoStop}%
\bibitem [{\citenamefont {Alvarez-Gaume}\ and\ \citenamefont
  {Ginsparg}(1985)}]{alvarez1985structure}%
  \BibitemOpen
  \bibfield  {author} {\bibinfo {author} {\bibfnamefont {L.}~\bibnamefont
  {Alvarez-Gaume}}\ and\ \bibinfo {author} {\bibfnamefont {P.}~\bibnamefont
  {Ginsparg}},\ }\href@noop {} {\bibfield  {journal} {\bibinfo  {journal}
  {Annals of Physics}\ }\textbf {\bibinfo {volume} {161}},\ \bibinfo {pages}
  {423} (\bibinfo {year} {1985})}\BibitemShut {NoStop}%
\bibitem [{\citenamefont {Ryu}\ \emph {et~al.}(2012)\citenamefont {Ryu},
  \citenamefont {Moore},\ and\ \citenamefont {Ludwig}}]{PhysRevB.85.045104}%
  \BibitemOpen
  \bibfield  {author} {\bibinfo {author} {\bibfnamefont {S.}~\bibnamefont
  {Ryu}}, \bibinfo {author} {\bibfnamefont {J.~E.}\ \bibnamefont {Moore}}, \
  and\ \bibinfo {author} {\bibfnamefont {A.~W.~W.}\ \bibnamefont {Ludwig}},\
  }\href {\doibase 10.1103/PhysRevB.85.045104} {\bibfield  {journal} {\bibinfo
  {journal} {Phys. Rev. B}\ }\textbf {\bibinfo {volume} {85}},\ \bibinfo
  {pages} {045104} (\bibinfo {year} {2012})}\BibitemShut {NoStop}%
\bibitem [{\citenamefont {Wang}\ \emph {et~al.}(2011)\citenamefont {Wang},
  \citenamefont {Qi},\ and\ \citenamefont {Zhang}}]{PhysRevB.84.014527}%
  \BibitemOpen
  \bibfield  {author} {\bibinfo {author} {\bibfnamefont {Z.}~\bibnamefont
  {Wang}}, \bibinfo {author} {\bibfnamefont {X.-L.}\ \bibnamefont {Qi}}, \ and\
  \bibinfo {author} {\bibfnamefont {S.-C.}\ \bibnamefont {Zhang}},\ }\href
  {\doibase 10.1103/PhysRevB.84.014527} {\bibfield  {journal} {\bibinfo
  {journal} {Phys. Rev. B}\ }\textbf {\bibinfo {volume} {84}},\ \bibinfo
  {pages} {014527} (\bibinfo {year} {2011})}\BibitemShut {NoStop}%
\bibitem [{\citenamefont {Stone}(2012)}]{PhysRevB.85.184503}%
  \BibitemOpen
  \bibfield  {author} {\bibinfo {author} {\bibfnamefont {M.}~\bibnamefont
  {Stone}},\ }\href {\doibase 10.1103/PhysRevB.85.184503} {\bibfield  {journal}
  {\bibinfo  {journal} {Phys. Rev. B}\ }\textbf {\bibinfo {volume} {85}},\
  \bibinfo {pages} {184503} (\bibinfo {year} {2012})}\BibitemShut {NoStop}%
\bibitem [{\citenamefont {Ringel}\ and\ \citenamefont
  {Stern}(2013)}]{PhysRevB.88.115307}%
  \BibitemOpen
  \bibfield  {author} {\bibinfo {author} {\bibfnamefont {Z.}~\bibnamefont
  {Ringel}}\ and\ \bibinfo {author} {\bibfnamefont {A.}~\bibnamefont {Stern}},\
  }\href {\doibase 10.1103/PhysRevB.88.115307} {\bibfield  {journal} {\bibinfo
  {journal} {Phys. Rev. B}\ }\textbf {\bibinfo {volume} {88}},\ \bibinfo
  {pages} {115307} (\bibinfo {year} {2013})}\BibitemShut {NoStop}%
\bibitem [{\citenamefont {Hosur}\ and\ \citenamefont
  {Qi}(2013)}]{hosur2013recent}%
  \BibitemOpen
  \bibfield  {author} {\bibinfo {author} {\bibfnamefont {P.}~\bibnamefont
  {Hosur}}\ and\ \bibinfo {author} {\bibfnamefont {X.}~\bibnamefont {Qi}},\
  }\href@noop {} {\bibfield  {journal} {\bibinfo  {journal} {Comptes Rendus
  Physique}\ } (\bibinfo {year} {2013})}\BibitemShut {NoStop}%
\bibitem [{\citenamefont {Burkov}\ and\ \citenamefont
  {Balents}(2011)}]{Burkov:2011zr}%
  \BibitemOpen
  \bibfield  {author} {\bibinfo {author} {\bibfnamefont {A.}~\bibnamefont
  {Burkov}}\ and\ \bibinfo {author} {\bibfnamefont {L.}~\bibnamefont
  {Balents}},\ }\href {http://arxiv.org/abs/1105.5138} {\bibfield  {journal}
  {\bibinfo  {journal} {Phys. Rev. Lett.}\ }\textbf {\bibinfo {volume} {107}},\
  \bibinfo {pages} {127205} (\bibinfo {year} {2011})},\ \Eprint
  {http://arxiv.org/abs/1105.5138} {1105.5138} \BibitemShut {NoStop}%
\bibitem [{\citenamefont {Cho}(2011)}]{cho2011possible}%
  \BibitemOpen
  \bibfield  {author} {\bibinfo {author} {\bibfnamefont {G.~Y.}\ \bibnamefont
  {Cho}},\ }\href@noop {} {\bibfield  {journal} {\bibinfo  {journal} {arXiv
  preprint arXiv:1110.1939}\ } (\bibinfo {year} {2011})}\BibitemShut {NoStop}%
\bibitem [{\citenamefont {Liu}\ \emph {et~al.}(2013)\citenamefont {Liu},
  \citenamefont {Ye},\ and\ \citenamefont {Qi}}]{Liu:2012ly}%
  \BibitemOpen
  \bibfield  {author} {\bibinfo {author} {\bibfnamefont {C.-X.}\ \bibnamefont
  {Liu}}, \bibinfo {author} {\bibfnamefont {P.}~\bibnamefont {Ye}}, \ and\
  \bibinfo {author} {\bibfnamefont {X.-L.}\ \bibnamefont {Qi}},\ }\href
  {http://arxiv.org/abs/1204.6551} {\bibfield  {journal} {\bibinfo  {journal}
  {Phys. Rev. B}\ }\textbf {\bibinfo {volume} {87}},\ \bibinfo {pages} {235306}
  (\bibinfo {year} {2013})},\ \Eprint {http://arxiv.org/abs/1204.6551}
  {1204.6551} \BibitemShut {NoStop}%
\bibitem [{\citenamefont {Chang}\ \emph {et~al.}(2013)\citenamefont {Chang},
  \citenamefont {Zhang}, \citenamefont {Liu}, \citenamefont {Zhang},
  \citenamefont {Feng}, \citenamefont {Li}, \citenamefont {Wang}, \citenamefont
  {Chen}, \citenamefont {Dai}, \citenamefont {Fang}, \citenamefont {Qi},
  \citenamefont {Zhang}, \citenamefont {Wang}, \citenamefont {He},
  \citenamefont {Ma},\ and\ \citenamefont {Xue}}]{ADMA:ADMA201203493}%
  \BibitemOpen
  \bibfield  {author} {\bibinfo {author} {\bibfnamefont {C.-Z.}\ \bibnamefont
  {Chang}}, \bibinfo {author} {\bibfnamefont {J.}~\bibnamefont {Zhang}},
  \bibinfo {author} {\bibfnamefont {M.}~\bibnamefont {Liu}}, \bibinfo {author}
  {\bibfnamefont {Z.}~\bibnamefont {Zhang}}, \bibinfo {author} {\bibfnamefont
  {X.}~\bibnamefont {Feng}}, \bibinfo {author} {\bibfnamefont {K.}~\bibnamefont
  {Li}}, \bibinfo {author} {\bibfnamefont {L.-L.}\ \bibnamefont {Wang}},
  \bibinfo {author} {\bibfnamefont {X.}~\bibnamefont {Chen}}, \bibinfo {author}
  {\bibfnamefont {X.}~\bibnamefont {Dai}}, \bibinfo {author} {\bibfnamefont
  {Z.}~\bibnamefont {Fang}}, \bibinfo {author} {\bibfnamefont {X.-L.}\
  \bibnamefont {Qi}}, \bibinfo {author} {\bibfnamefont {S.-C.}\ \bibnamefont
  {Zhang}}, \bibinfo {author} {\bibfnamefont {Y.}~\bibnamefont {Wang}},
  \bibinfo {author} {\bibfnamefont {K.}~\bibnamefont {He}}, \bibinfo {author}
  {\bibfnamefont {X.-C.}\ \bibnamefont {Ma}}, \ and\ \bibinfo {author}
  {\bibfnamefont {Q.-K.}\ \bibnamefont {Xue}},\ }\href {\doibase
  10.1002/adma.201203493} {\bibfield  {journal} {\bibinfo  {journal} {Advanced
  Materials}\ }\textbf {\bibinfo {volume} {25}},\ \bibinfo {pages} {1065}
  (\bibinfo {year} {2013})}\BibitemShut {NoStop}%
\bibitem [{\citenamefont {Srednicki}(2007)}]{Srednicki:2007quantum}%
  \BibitemOpen
  \bibfield  {author} {\bibinfo {author} {\bibfnamefont {M.}~\bibnamefont
  {Srednicki}},\ }\href@noop {} {\emph {\bibinfo {title} {{Quantum Field
  Theory}}}}\ (\bibinfo  {publisher} {Cambridge University Press},\ \bibinfo
  {year} {2007})\BibitemShut {NoStop}%
\bibitem [{Note1()}]{Note1}%
  \BibitemOpen
  \bibinfo {note} {When $f_{\mu \nu } =0$, the solution to Dirac equation is
  given by $\psi =0$. In this letter, we would like to obtain the effective
  interaction between magnons and electromagnetic fields by integrating out
  Weyl fermion fluctuations around the vacuum solution $\psi =0$. However, when
  the flux of the axial vector field strength takes non-zero integer values,
  the solution to Dirac equation consists of additional $(1+1)$-dimensional
  Weyl fermions \cite {Liu:2012ly}. The topological response obtained by
  integrating out fermionic fluctuations around this non-trivial background
  will be studied elsewhere.}\BibitemShut {Stop}%
\bibitem [{\citenamefont {Qi}\ \emph {et~al.}(2008)\citenamefont {Qi},
  \citenamefont {Hughes},\ and\ \citenamefont {Zhang}}]{PhysRevB.78.195424}%
  \BibitemOpen
  \bibfield  {author} {\bibinfo {author} {\bibfnamefont {X.-L.}\ \bibnamefont
  {Qi}}, \bibinfo {author} {\bibfnamefont {T.~L.}\ \bibnamefont {Hughes}}, \
  and\ \bibinfo {author} {\bibfnamefont {S.-C.}\ \bibnamefont {Zhang}},\ }\href
  {\doibase 10.1103/PhysRevB.78.195424} {\bibfield  {journal} {\bibinfo
  {journal} {Phys. Rev. B}\ }\textbf {\bibinfo {volume} {78}},\ \bibinfo
  {pages} {195424} (\bibinfo {year} {2008})}\BibitemShut {NoStop}%
\bibitem [{\citenamefont {Adler}(1969)}]{PhysRev.177.2426}%
  \BibitemOpen
  \bibfield  {author} {\bibinfo {author} {\bibfnamefont {S.~L.}\ \bibnamefont
  {Adler}},\ }\href {\doibase 10.1103/PhysRev.177.2426} {\bibfield  {journal}
  {\bibinfo  {journal} {Phys. Rev.}\ }\textbf {\bibinfo {volume} {177}},\
  \bibinfo {pages} {2426} (\bibinfo {year} {1969})}\BibitemShut {NoStop}%
\bibitem [{\citenamefont {Bell}\ and\ \citenamefont
  {Jackiw}(1969)}]{bell1969pcac}%
  \BibitemOpen
  \bibfield  {author} {\bibinfo {author} {\bibfnamefont {J.~S.}\ \bibnamefont
  {Bell}}\ and\ \bibinfo {author} {\bibfnamefont {R.}~\bibnamefont {Jackiw}},\
  }\href@noop {} {\bibfield  {journal} {\bibinfo  {journal} {Il Nuovo Cimento
  A}\ }\textbf {\bibinfo {volume} {60}},\ \bibinfo {pages} {47} (\bibinfo
  {year} {1969})}\BibitemShut {NoStop}%
\bibitem [{\citenamefont {t~Hooft}\ and\ \citenamefont
  {Veltman}(1972)}]{t1972regularization}%
  \BibitemOpen
  \bibfield  {author} {\bibinfo {author} {\bibfnamefont {G.}~\bibnamefont
  {t~Hooft}}\ and\ \bibinfo {author} {\bibfnamefont {M.}~\bibnamefont
  {Veltman}},\ }\href@noop {} {\bibfield  {journal} {\bibinfo  {journal}
  {Nuclear Physics B}\ }\textbf {\bibinfo {volume} {44}},\ \bibinfo {pages}
  {189} (\bibinfo {year} {1972})}\BibitemShut {NoStop}%
\bibitem [{\citenamefont {Gottlieb}\ and\ \citenamefont
  {Donohue}(1979)}]{PhysRevD.20.3378}%
  \BibitemOpen
  \bibfield  {author} {\bibinfo {author} {\bibfnamefont {S.}~\bibnamefont
  {Gottlieb}}\ and\ \bibinfo {author} {\bibfnamefont {J.~T.}\ \bibnamefont
  {Donohue}},\ }\href {\doibase 10.1103/PhysRevD.20.3378} {\bibfield  {journal}
  {\bibinfo  {journal} {Phys. Rev. D}\ }\textbf {\bibinfo {volume} {20}},\
  \bibinfo {pages} {3378} (\bibinfo {year} {1979})}\BibitemShut {NoStop}%
\bibitem [{\citenamefont
  {Ho{\v{r}}ej{\v{s}}{\'\i}}(1992)}]{hovrejvsi1992ultraviolet}%
  \BibitemOpen
  \bibfield  {author} {\bibinfo {author} {\bibfnamefont {J.}~\bibnamefont
  {Ho{\v{r}}ej{\v{s}}{\'\i}}},\ }\href@noop {} {\bibfield  {journal} {\bibinfo
  {journal} {Czech. J. Phys.}\ }\textbf {\bibinfo {volume} {42}},\ \bibinfo
  {pages} {345} (\bibinfo {year} {1992})}\BibitemShut {NoStop}%
\bibitem [{\citenamefont {Harvey}()}]{Harvey:fk}%
  \BibitemOpen
  \bibfield  {author} {\bibinfo {author} {\bibfnamefont {J.~A.}\ \bibnamefont
  {Harvey}},\ }\href {http://arxiv.org/abs/hep-th/0509097} {\ }\Eprint
  {http://arxiv.org/abs/hep-th/0509097} {hep-th/0509097} \BibitemShut {NoStop}%
\bibitem [{\citenamefont {Zyuzin}\ and\ \citenamefont
  {Burkov}(2012)}]{PhysRevB.86.115133}%
  \BibitemOpen
  \bibfield  {author} {\bibinfo {author} {\bibfnamefont {A.~A.}\ \bibnamefont
  {Zyuzin}}\ and\ \bibinfo {author} {\bibfnamefont {A.~A.}\ \bibnamefont
  {Burkov}},\ }\href {\doibase 10.1103/PhysRevB.86.115133} {\bibfield
  {journal} {\bibinfo  {journal} {Phys. Rev. B}\ }\textbf {\bibinfo {volume}
  {86}},\ \bibinfo {pages} {115133} (\bibinfo {year} {2012})}\BibitemShut
  {NoStop}%
\bibitem [{\citenamefont {Wang}\ and\ \citenamefont
  {Zhang}(2013)}]{PhysRevB.87.161107}%
  \BibitemOpen
  \bibfield  {author} {\bibinfo {author} {\bibfnamefont {Z.}~\bibnamefont
  {Wang}}\ and\ \bibinfo {author} {\bibfnamefont {S.-C.}\ \bibnamefont
  {Zhang}},\ }\href {\doibase 10.1103/PhysRevB.87.161107} {\bibfield  {journal}
  {\bibinfo  {journal} {Phys. Rev. B}\ }\textbf {\bibinfo {volume} {87}},\
  \bibinfo {pages} {161107} (\bibinfo {year} {2013})}\BibitemShut {NoStop}%
\bibitem [{\citenamefont {Yang}\ \emph {et~al.}(2011)\citenamefont {Yang},
  \citenamefont {Lu},\ and\ \citenamefont {Ran}}]{PhysRevB.84.075129}%
  \BibitemOpen
  \bibfield  {author} {\bibinfo {author} {\bibfnamefont {K.-Y.}\ \bibnamefont
  {Yang}}, \bibinfo {author} {\bibfnamefont {Y.-M.}\ \bibnamefont {Lu}}, \ and\
  \bibinfo {author} {\bibfnamefont {Y.}~\bibnamefont {Ran}},\ }\href {\doibase
  10.1103/PhysRevB.84.075129} {\bibfield  {journal} {\bibinfo  {journal} {Phys.
  Rev. B}\ }\textbf {\bibinfo {volume} {84}},\ \bibinfo {pages} {075129}
  (\bibinfo {year} {2011})}\BibitemShut {NoStop}%
\bibitem [{\citenamefont {Chen}\ \emph {et~al.}(2013)\citenamefont {Chen},
  \citenamefont {Wu},\ and\ \citenamefont {Burkov}}]{PhysRevB.88.125105}%
  \BibitemOpen
  \bibfield  {author} {\bibinfo {author} {\bibfnamefont {Y.}~\bibnamefont
  {Chen}}, \bibinfo {author} {\bibfnamefont {S.}~\bibnamefont {Wu}}, \ and\
  \bibinfo {author} {\bibfnamefont {A.~A.}\ \bibnamefont {Burkov}},\ }\href
  {\doibase 10.1103/PhysRevB.88.125105} {\bibfield  {journal} {\bibinfo
  {journal} {Phys. Rev. B}\ }\textbf {\bibinfo {volume} {88}},\ \bibinfo
  {pages} {125105} (\bibinfo {year} {2013})}\BibitemShut {NoStop}%
\bibitem [{\citenamefont {Mahan}(2000)}]{Mahan}%
  \BibitemOpen
  \bibfield  {author} {\bibinfo {author} {\bibfnamefont {G.~D.}\ \bibnamefont
  {Mahan}},\ }\href@noop {} {\emph {\bibinfo {title} {Many-Particle
  Physics}}},\ \bibinfo {edition} {3rd}\ ed.\ (\bibinfo  {publisher}
  {Springer},\ \bibinfo {year} {2000})\BibitemShut {NoStop}%
\bibitem [{\citenamefont {Mahan}(2010)}]{mahan2010condensed}%
  \BibitemOpen
  \bibfield  {author} {\bibinfo {author} {\bibfnamefont {G.~D.}\ \bibnamefont
  {Mahan}},\ }\href@noop {} {\emph {\bibinfo {title} {Condensed matter in a
  nutshell}}}\ (\bibinfo  {publisher} {Princeton University Press},\ \bibinfo
  {year} {2010})\BibitemShut {NoStop}%
\bibitem [{\citenamefont {Li}\ \emph {et~al.}(2010)\citenamefont {Li},
  \citenamefont {Wang}, \citenamefont {Qi},\ and\ \citenamefont
  {Zhang}}]{li2010dynamical}%
  \BibitemOpen
  \bibfield  {author} {\bibinfo {author} {\bibfnamefont {R.}~\bibnamefont
  {Li}}, \bibinfo {author} {\bibfnamefont {J.}~\bibnamefont {Wang}}, \bibinfo
  {author} {\bibfnamefont {X.-L.}\ \bibnamefont {Qi}}, \ and\ \bibinfo {author}
  {\bibfnamefont {S.-C.}\ \bibnamefont {Zhang}},\ }\href@noop {} {\bibfield
  {journal} {\bibinfo  {journal} {Nature Physics}\ }\textbf {\bibinfo {volume}
  {6}},\ \bibinfo {pages} {284} (\bibinfo {year} {2010})}\BibitemShut {NoStop}%
\bibitem [{\citenamefont {Popova}\ \emph {et~al.}(2001)\citenamefont {Popova},
  \citenamefont {Keller}, \citenamefont {Gendron}, \citenamefont {Guyot},
  \citenamefont {Brianso}, \citenamefont {Dumond},\ and\ \citenamefont
  {Tessier}}]{popova2001structure}%
  \BibitemOpen
  \bibfield  {author} {\bibinfo {author} {\bibfnamefont {E.}~\bibnamefont
  {Popova}}, \bibinfo {author} {\bibfnamefont {N.}~\bibnamefont {Keller}},
  \bibinfo {author} {\bibfnamefont {F.}~\bibnamefont {Gendron}}, \bibinfo
  {author} {\bibfnamefont {M.}~\bibnamefont {Guyot}}, \bibinfo {author}
  {\bibfnamefont {M.-C.}\ \bibnamefont {Brianso}}, \bibinfo {author}
  {\bibfnamefont {Y.}~\bibnamefont {Dumond}}, \ and\ \bibinfo {author}
  {\bibfnamefont {M.}~\bibnamefont {Tessier}},\ }\href@noop {} {\bibfield
  {journal} {\bibinfo  {journal} {Journal of Applied Physics}\ }\textbf
  {\bibinfo {volume} {90}},\ \bibinfo {pages} {1422} (\bibinfo {year}
  {2001})}\BibitemShut {NoStop}%
\end{thebibliography}%

\end{document}